\documentclass[12pt,preprint]{aastex}
\usepackage{lscape}

\def\Mu5{\dot{\cal{M}}}

\begin{document}

\title{X-rays from the explosion site: Fifteen years of light curves of SN 1993J}

\author{Poonam Chandra\altaffilmark{1,2},
Vikram V. Dwarkadas\altaffilmark{3}
Alak Ray\altaffilmark{4},
Stefan Immler\altaffilmark{5},
\&
David Pooley\altaffilmark{6},
}
\altaffiltext{1}{Jansky Fellow, National Radio Astronomy Observatory}
\altaffiltext{2}{Department of Astronomy, University of Virginia, P.O. Box
        400325, Charlottesville, VA 22904}
\altaffiltext{3}{Department of Astronomy and Astrophysics, 
University of Chicago, 5640 S Ellis Ave, AAC 010c,  
Chicago, Illinois 60637}
\altaffiltext{4}{Tata Institute of Fundamental Research, Mumbai 400 005, India}
\altaffiltext{5}{NASA GSFC, Greenbelt, MD 20771}
\altaffiltext{6}{University of Wisconsin,
4512 Sterling Hall, Madison, WI 53706}
\email{pc8s@virginia.edu}

\slugcomment{Submitted to \textit{ApJ}}

\shorttitle{X-ray light-curves of SN 1993J}
\shortauthors{Chandra \textit{et al.}}

\begin{abstract}
We present a comprehensive analysis of the X-ray light curves of SN
1993J in a nearby galaxy M81.  This is the only supernova other than
SN 1987A, which is so extensively followed in the X-ray bands.  
  Here we report on SN 1993J observations with the {\it Chandra} in
  the year 2005 and 2008, and Swift observations in 2005, 2006 and 2008. We
  combined these observations with all available archival data of SN 1993J,
  which includes ROSAT, ASCA, {\it Chandra}, and XMM-{\it Newton}
  observations from 1993 April to 2006 August.  In this paper we
  report the X-ray light curves of SN 1993J, extending up to
fifteen years,  in the soft
  (0.3--2.4~keV), hard (2--8~keV) and combined (0.3--8~keV) bands.
  The hard and soft-band fluxes decline at different rates initially,
  but after about 5 years they both undergo a $t^{-1}$ decline.  The
  soft X-rays, which are initially low, start dominating after a few
  hundred days.  
We interpret that most of the emission below 8~keV is coming from
  the reverse shock which is radiative initially for around first
  1000-2000 days and then turn into adiabatic shock. 
Our hydrodynamic
  simulation also confirms the reverse shock origin of the observed
  light curves.  We also compare the
H$\alpha$ line luminosity of SN 1993J with its
 X-ray light curve and note that
the H$\alpha$ line luminosity has
a fairly high fraction of the X-ray emission, indicating 
presence of clumps in the emitting plasma. 

\end{abstract}


\section{Introduction}
\label{sec:introduction}

The explosion of a massive star as a supernova (SN) can
drive powerful shocks into the circumstellar medium (CSM) of the
progenitor.  The CSM is established by the mass lost from the
progenitor prior to explosion.  Collision of the ejected material with
the CSM leads to a blast wave shock with velocities $10,000-20,000$
km/s and a hot shell of $T \sim 10^9 K$.  The interface of the
expanding ejecta where it first meets the circumstellar gas is the
contact discontinuity which itself propagates outwards with time. As
the ejecta expands, the interface region between the blast wave shock
and the contact discontinuity sweeps up matter from the circumstellar
gas and decelerates due to the accumulated mass. This in turn 
gives rise to a shocked shell that propagates back into the cold expanding ejecta
This 'reverse'
shock propagates inwards at $\sim 10^3$ km/s, significantly slower
speed than the fastest {\it expanding} stellar ejecta and has a temperature
of $T \sim 10^7 $ K.  The reverse shock is believed to be the site of
most of the observable X-ray emission at late times.

In this paper, we describe the long term light curves of the SN 1993J
in various X-ray bands. We briefly describe in \S \ref{sec:1993j}
the current understanding about the unusual nature of SN 1993J and its
progenitor star as obtained from multi-waveband observations and
theoretical analysis. We discuss the X-ray emission mechanisms and the
previous X-ray studies of SN 1993J in \S \ref{sec:previous}. In \S
\ref{sec:observations}, we discuss the X-ray observations and analysis
of SN 1993J over the years made with several X-ray telescopes.  Our
detailed results and interpretation are mentioned in \S
\ref{sec:results}.  Hydrodynamical simulations are discussed in \S
\ref{sec:hydro}, and X-ray versus optical H$\alpha$ correlation 
in \S
\ref{sec:discussion}.  We report our main conclusions in \S
\ref{sec:conclusion}.

\section{A supernova that underwent a metamorphosis of spectral types}
\label{sec:1993j}

SN 1993J is one of the best studied supernovae (SNe) in all wavelength regimes, 
second only to SN 1987A.
It was visually discovered on 1993 March 28.906 UT
\citep{rip93},  
in the nearby galaxy M81 \citep[aka NGC 3031 at
$d = 3.63~{\rm Mpc}$,][]{fre94}. It was the 
optically brightest SN in the northern
hemisphere since SN 1954A, having reached a secondary maximum brightness of
$V = 10.8$ mag on day 21.1.

SN 1993J is a SN that has undergone an ``identity crisis".  It's
spectrum
underwent a transition from a type II spectrum (characterized by
strong hydrogen Balmer lines) at early epochs, to a type Ib-like
spectrum at $\approx 300$ days (with its nebular spectra having weak
hydrogen but strong He I lines).  This SN was classified as
Type IIb SN and provided for the first time a link between type
II supernovae (SNe) and type Ib SNe \citep{fil93,swa93}.  Models of SN 1993J based
on the early light curve indicated that the progenitor star lost all
but a small amount of its hydrogen layers due to mass transfer in a
binary system \citep{nom93,ray93,pod93,woo94}.  Thus the shock heated
photosphere could quickly recede through the small H layer into the
deeper He layers during the initial expansion and cooling phase
itself. Among core collapse SNe, it is thus possible to have a
continuum of hydrogen envelope masses remaining on the progenitor star when it
explodes.

At early epochs SN 1993J showed the typical signatures of
circumstellar (CS) interaction in the radio \citep{van94}, UV \citep{fra94}
and X-ray \citep{zim94} wavelengths.  The UV and optical spectra taken
with the Hubble Space Telescope (HST) and Keck telescopes have
revealed the signature of a massive star, along with the fading
supernova 10 years after explosion, which is the binary companion of
the progenitor that exploded \citep{mau04}.  The circumstellar medium
has therefore been modified by the mass loss in a binary system in SN
1993J.

\section{X-ray emission and early X-ray studies of SN 1993J}
\label{sec:previous}

The X-ray luminosity and its time variation depends, among other
things, on the density structure of the stellar ejecta.  In core
collapse SNe, the explosion dynamics quickly leads to an ejecta
outer density profile with a steep power-law 
( $\rho_{\rm ej}\propto r^{-n}$ where $n$ is a constant) 
in the radial
coordinate, but with a relatively flat inner core density profile
\citep{che89, mat99}.  
The ejecta density
profile depends on the initial structure of the star and, in particular,
is affected by if the progenitor had a radiative or a convective
envelope \citep{mat99}.  The shock propagation through the outer
profile does not depend upon the behaviour of the inner layers and a
limiting structure is described by a self-similar solution as the
shock front accelerates while propagating through the outer layers
with rapidly decreasing density.  The self similar nature of the
evolution of the ejecta dominated SNe (or supernova remnants (SNRs))
 applies only for
ejecta with steep envelope index $n > 5$, since for such steep $n$,
the mass and energy of the ejecta remain finite \citep{tru99}.

 The shocked shells (forward or reverse) generated due to ejecta-wind
 interaction have very high temperatures (\S \ref{sec:introduction})
 and can emit X-rays.  The forward shock is adiabatic almost all the
 time and its luminosity follows the time dependence of $L_{\rm
   forward} \propto t^{-1}$ \citep{che03}.  However, depending upon
 the ejecta density profile and the mass loss rate of the SN
 progenitor star, the reverse shock can be either adiabatic or
 radiative, or can be radiative at early phase and then make a
 transition to adiabatic phase.
The expression for free-free emission from 
adiabatic reverse shock and its luminosity
is derived by  \cite{che03} and \citet{fra96}:
\begin{equation}
L_{\rm rev}^{\rm ad} \approx 3\times 10^{40} \bar g_{\rm ff}
\frac{(n-3)(n-4)^2}{4(n-2)}
\left( {\dot M_{-5}  \over u_{\rm w1}}\right)^2
\left( t \over 1\,\rm day \right)^{-1}\,\, \rm erg\,s^{-1}.
\label{eq:adia}
\end{equation}
Here $\dot M_{-5}$ is progenitor mass loss rate in
units of  $10 ^{-5} \rm
M_\odot \, yr^{-1}$, $u_{\rm w1}$ is the wind velocity in units of 10
km s$^{-1}$, and $\bar g_{\rm ff}$ is the free-free Gaunt factor. This
shows that free-free emission from adiabatic reverse shock follows
time evolution of $t^{-1}$ for constant mass-loss rate and wind
velocity. However, if line emission dominates, the time dependence
of X-ray line luminosity is
 $t^{1.2-2.2m}$ \citep{che94}, where $m$ is the expansion parameter
in $R \propto t^{m}$.

\citet{fra96} have discussed that in the initial phase, reverse shock
with high density gradient is likely to be radiative resulting
in formation of a cooled shell between the reverse shock and the
forward shock.  When the electron temperature is $T_e \leq 2 \times
10^7$ K, the line emission dominates the total X-ray emission.
\citet{che03} have discussed the importance of line emission and its
effect on the cooling rate of the gas behind the reverse shock.  In
case of line emission, a thermal instability develops and the gas cools
up to about $10^4$ K where the temperature is stabilized only by
photoelectric heating from the shock balancing the cooling. For an
ejecta velocity scale of $V_{ej}$, and a reverse shock moving through
an ejecta density gradient $\rho \propto r^{-n}$ with solar
composition one obtains a cooling time scale from $t_{\rm cool}
=3kT_{\rm e}/n \Lambda$, where $\Lambda$ is the cooling 
function.  The cooling time  
can be expanded as 
\citep{fra96,
  che03}:
\begin{equation}
t_{\rm cool} =  {605 \over  (n-3) (n-4) (n -
2)^{3.34}}\left({V_{ej} \over 10^4 \, \rm km \,s^{-1}}\right)^{5.34} 
\left({\dot M_{-5} \over
u_{\rm w 1}}\right)^{-1} \left({t\over {1\,\rm  day}}\right)^2~~{\rm days},
\label{eq:tcool}
\end{equation}
Since  large exponents of the velocity scale and density gradient index are
involved, it is clear that the cooling time is sensitive to them, and also
the mass loss rate of the pre-explosion progenitor star. 
The most important effect of cooling gas
 between the reverse shock and
the observer is that the cool gas absorbs most 
of the emission from the reverse
shock, and in spite of the higher intrinsic
 luminosity of the reverse shock,
little of it will be directly observable
 initially. The column density of the
cool gas also thins out as the supernova ages and expands in scale. 
In such a situation the total luminosity of the reverse shock
may contribute appreciably, or even dominate, to 
the bolometric luminosity and is defined as:
\begin{eqnarray}
L_{\rm rev}^{\rm rad} = 
 1.6\times 10^{41} ~ {(n-3)(n-4)\over
(n-2)^3} \left( \dot M_{-5} \over u_{\rm w1} \right) 
\left(V_{rev} \over 10^4 \, \rm km \,s^{-1} \right)^3\, \rm erg \; s^{-1}
\label{eq:rad}
\end{eqnarray}
Since $V_{\rm rev} \propto t^{-1/(n-2)}$, the 
time dependence of luminosity in cooling case is
$L_{\rm rev}^{\rm rad} \propto t^{-3/(n-2)}$. In Table \ref{tab:tcool},
we tabulate the the timescales up to which the reverse shock remains
radiative under different conditions. We take the case with
density index of 7, 12 and 20 for three compositions: solar, helium
and oxygen. As one can see from the table, the deeper in the
ejecta the reverse shock (where the composition is dominated by heavier elements)
is, the longer time do the reverse shock remain radiative.

It has been argued
that SN 1993J like objects with their high mass loss rate will have 
radiative reverse shocks (where cooling is important) even at
late epochs ($\geq$ 100 days)
whereas SNe with low mass loss rates such as SN 1999em (a type IIP SN)  with
long cooling times will develop adiabatic
 reverse shocks early
on \citep{nym06}. However in some cases 
adiabatic and radiative reverse shocks co-exist as 
has been seen in SN 1987A \citep{gro06}.
\citet{nym09} has discussed that if the ejecta or the CSM is clumpy, the
adiabatic reverse shock may give rise to slow moving oblique
shocks which are radiative in nature.

Soon after its discovery SN 1993J was observed and detected in the
X-rays by the ROSAT \citep{zim93}. Since then, the supernova has been
observed at various epochs with a number of X-ray telescopes, including ROSAT
\citep{zim94,imm01}, ASCA \citep{tan93,koh94,uno02}, {\it Chandra}
\citep[][and this paper]{swa03}, XMM-Newton \citep{zim03} and {\it
  Swift} (this paper) satellite missions.  The Oriented Scintillation
Spectrometer Experiment (OSSE) on the Compton Observatory had detected
the SN in 50--150~keV energy band on day 12 and 30 
with very high X-ray luminosity ($5 \times 10^{40}$ erg
s$^{-1}$) on day 12 \citep{lei94}. However, they found that the 50--150~keV 
emission faded below detection by day 108.

 \citet{suz93} did a detailed analysis of the early X-ray emission
  from the SN. They carried out hydrodynamical modelling of the
  collision between the ejecta and a CSM created by steady winds
  ($\rho_{CSM} \propto r^{-2}$), and claimed that the observed
  features of X-ray emission can be accounted for with thermal
  free-free emission from the shock heated ejecta. They also
  predicted that the expansion velocities of the ejecta must be high
  and the density gradient of the ejecta shallow.  Later they extended
  their analysis to X-ray observations of first 600 days
in \citet{suz95}. In this paper they assumed a more realistic ejecta
model, a steep density gradient in the outermost layer followed by
a much shallow density gradient in the inner H-rich envelope. The
density profile suddenly increases and becomes steeper again at the
interface of H-rich envelope and the He-core. The CSM was assumed to
be clumpy and cooling effects were taken into account with
formation of a dense
cooling shell between the reverse and the forward shock. Early X-rays
from the reverse shock were absorbed in this cool shell and X-rays
were mainly the thermal emission from the shocked CSM.  Their results
indicated that the CSM has density distribution of the nature
$\rho_{CSM} \propto r^{-1.7}$ instead of $\rho_{CSM} \propto r^{-2}$,
indicating that the CSM was created by non-steady winds.  
According to their model, the CSM has to be more clumpy in the
outer layers to explain the observed X-ray features. The collisions
between the forward shock and the clumps may give rise to soft
X-rays. Such collisions would take place with many clumps at various
distances, and the sum of resulting X-ray emissions may form a more or
less continuous light curve.  In their model, all the soft X-rays in
first few tens of days were coming from the clumpy CSM, and had no
contribution from the reverse shell as they were totally absorbed by
the cooled shell. However, in \citet{nom98}, they revised their
prediction and claimed that the cooling dense shell formed between the
reverse and the forward shocked shell is clumpy due to the
Rayleigh-Taylor instability and thus some early X-rays may also leak
out from reverse shock. They also predicted the possible increase in
the soft X-rays around day 500 to be due to the density jump from the
H-rich envelope to He-layer, modified by the Rayleigh-Taylor
instabilities.

\citet{fra96} have studied the early time X-ray emission from SN 1993J
in detail.  Using radio and X-ray observations from the first two
weeks, they estimated the mass loss rate of $\sim 4 \times 10^{-5} \rm
M_\odot \, yr^{-1} $. Fitting the early radio data using the free-free
absorption model, they suggested a wind density profile: 
 $\rho_{CSM} \propto r^{-1.5}$. They did the detailed
analysis of X-ray data from {\it ROSAT}, {\it ASCA} and OSSE on the
{\it Compton Observatory}. According to their conclusion, the early
hard X-rays arise mainly from the CS shock, whereas the high column
density of the cooling shell between the CS shock and the radiative
reverse shock absorbs most of the soft X-ray emission arising from the
radiative reverse shock. After about 50-100 days the column density of
the cooling shell decreases enough for the X-rays from the reverse
shock to leak out and dominate the spectrum. They also found that the
X-ray observations are best fit with the flatter CSM density profile
of $\rho_{CSM} \propto r^{-1.7}$.  However, in their later 
work \citet{fra98} claimed that
the CSM density profile is more sensitive to the radio observations
and they found the radio observations to be fitting well in
$\rho_{CSM} \propto r^{-2}$ profile.  The earlier discrepancy in radio
band was claimed to be due to neglecting the synchrotron-self absorption process
which plays an important role in the radio absorption for SN
1993J. 
 However, \citet{mio01}  included both
  synchrotron self-absorption and free-free absorption in their
  analysis and claimed that the radio emission could not be fit with a CSM
  declining as $r^{-2}$, but that a density profile going as
  $r^{-1.7}$ provided a much better fit. In this paper, we use the
CSM profile to be $\rho_{CSM} \propto r^{-2}$, though we explore the possibility
of different profiles in \S \ref{sec:hydro}. 

\section{Observations and Analysis} 
\label{sec:observations}

As mentioned in \S \ref{sec:previous}, SN 1993J has been observed
at various epochs with multiple X-ray telescopes since its discovery. We describe 
below these observations in detail. We analyzed  the 2005 and 2008 {\it Chandra}
data and the {\it Swift} data at several epochs between 2006-2008.
We also reanalyzed the 2001 XMM-Newton and 2000 {\it Chandra}
 datasets and  extracted 
unabsorbed fluxes in 0.3--2.4~keV, 2--8~keV and 0.3--8~keV
bands. For ROSAT and 
ASCA, we extracted unabsorbed fluxes in the above bands 
using best fit parameters mentioned in various 
references (see below). 
In our fits, the column density is a measure of the absorbing cool shell 
in the SN plus the
galactic absorption shell.
The column density was set to  zero to extract the absorption corrected 
luminosities. 
Below we describe the various SN 1993J X-ray observations in detail.
The details of all
 observations used in this paper are
are summarized in Table \ref{tab:obs}.

\subsection{ROSAT observations}

X-ray observations of SN 1993J by  {\it ROSAT} between days 6 to 1181 after
explosion was reported by \citet{imm01}, in which 
both PSPC and HRI observations were included. HRI data were
binned in observation blocks of length 5--20~ks exposure time
whereas the PSPC observation blocks had integration times ranging
2--18 ks long. They had assumed a constant absorption column density 
and extracted the fluxes. 
\citet{zim03} reanalyzed all the {\it ROSAT} data and reported the 
observations up to day 1800. They fit all the spectra 
assuming column density as a free parameter and with a thermal 
plasma model with a single temperature (vmekal in XSPEC) 
 fixing the element abundances
to  that obtained from, e.g.
{\it XMM-Newton} PN spectrum. They found that due to the
 high temperatures in the early ROSAT observations there is almost no difference
between fitting either with solar or with the {\it XMM-Newton} elemental
abundances. We use the best fit parameters quoted in Table 3 of \citet{zim03}
to extract the ROSAT fluxes of the SN at various epochs.
Table \ref{tab:1} shows
the unabsorbed luminosities for the ROSAT-HRI and ROSAT-PSPC 
observations in the energy range of 0.3--2.4~keV.

\subsection{ASCA observations}

ASCA flux points were extracted from \citet{koh94},
 \citet{uno02} and \citet{swa03}. 
The large Field of View (FoV) of ASCA 
showed  the presence of the SN host galaxy and
a bright X-ray binary source
at respectively
$3'$  and  $1'$ away from the SN. 
Since ASCA PSF has resolution of $3'$, 
the bright X-ray binary
was strongly contaminating the SN flux. To avoid this
contamination \citet{koh94} and \citet{uno02} isolated SN 1993J flux
through a one-dimensional image fitting, by 
using  $1' \times 2'$ rectangular boxes
containing the SN and the X-ray binary separately,  and by modelling the
one-dimensional intensity profile. They claimed to 
get rid of almost all the contamination from the M81 and the 
X-ray binary source, by this method. 

To estimate more reliable ASCA fluxes of SN 1993J, we attempted to utilize the 
late {\it Chandra} observations. Due to its high 
angular resolution, {\it Chandra} 
is able to easily separate the SN, and the nearby X-ray binary and M81
nucleus.  Our aim was to extract the uncontaminated flux of the X-ray
binary and then subtract it from all the ASCA observations. However, for
this technique to work, we needed to make sure that the X-ray binary
is not a time variable source.  To determine this, we analyzed two
{\it Chandra} datasets, one of 2000 May 7 and another on 2005 Jun 01
and extracted the flux of the X-ray binary from both the data
respectively.  Our analysis shows that the X-ray binary is a highly
variable source, especially in the hard X-rays. The $0.3-8.0$ keV flux
decreases by a factor of two in five years, from $2.79 \times 10^{-12}$ erg
cm$^{-2}$ s$^{-1}$ in May 2000 to $1.40 \times 10^{-12}$ erg cm$^{-2}$
s$^{-1}$ in June 2005. The change in the SN flux at $2-8$ keV is by
more than a factor of three in the two observations
 ($2.26 \times 10^{-12}$ erg cm$^{-2}$ s$^{-1}$
in May 2000 to $6.9 \times 10^{-13}$ erg cm$^{-2}$ s$^{-1}$ in June
2005).  Thus, in view of high variability of the X-ray binary source, it is
not advisable to reanalyze all the ASCA datasets to remove the
contribution of the X-ray binary on the basis of {\it Chandra} flux of
the binary. 
We use the spectral fit parameters of \citet{koh94}
and \citet{uno02} to convert the count rates into unabsorbed fluxes in
bands 0.3--2.4, 2--8 and 0.3--8~keV fluxes (Table \ref{tab:1}).

\subsection{{\it Chandra} observations}

{\it Chandra} first observed SN 1993J on
 2000 March 21 and 2000 May 07 under 
ObsIDs 390 and 735 with ACIS-S by  \citet{swa03}. 
They showed that
SN 1993J  faded since its discovery and its spectrum softened. 
At this time SN 1993J was $\sim 2600$ days old and displayed a complex thermal 
spectrum from a reverse shock rich in Fe L and highly ionized 
Mg, Si, and S but lacking Oxygen.
The unabsorbed luminosity reported in Table \ref{tab:1} are extracted
from \citet{swa03}.

\begin{figure}
\begin{center}
\includegraphics[angle=270,width=0.7\textwidth]{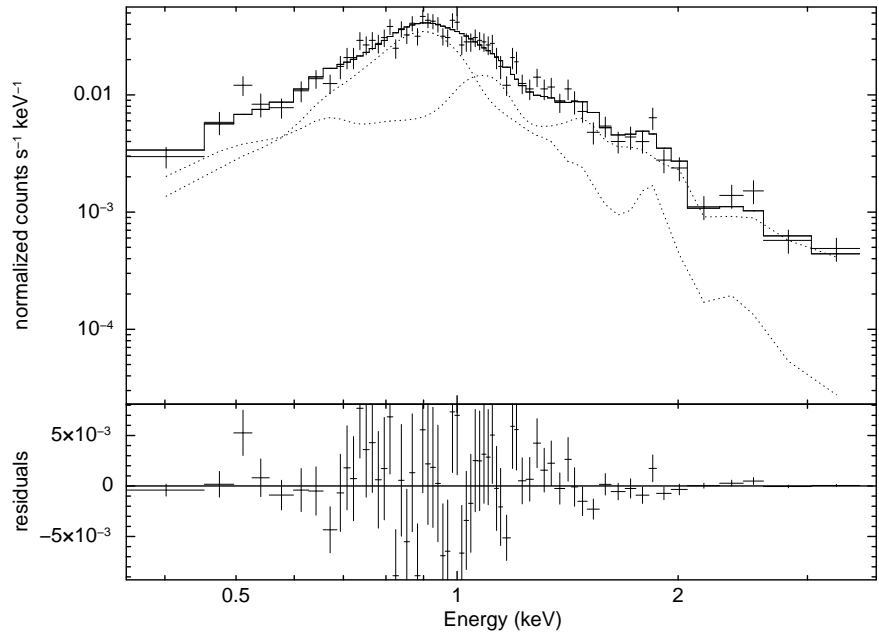}
\caption{Here we plot the
two component Mekal fit to SN 1993J {\it Chandra} data
of 2005 taken with ACIS-S.
Best fit temperatures from the Mekal model are 0.73 keV and
2.21 keV with absorption column density of $6 \times 10^{20}$ cm$^{-2}$.
}
\label{fig:spec}
\end{center}
\end{figure}

{\it Chandra} ACIS-S  observed the SN starting from May 26, 2005 to
July 6, 2005 on 15 different occasions (Obs Ids: 5935-5949; PI: Pooley).
Each observation was around 11-12 ks.
 We analyzed this data using CIAO analysis threads and XSPEC.
Event 2 files (pipeline processed files) were used for the data analysis.
Standard methods were used to analyze the data \citep{cha05}. 
We combined the whole data set in
two groups to increase the signal to noise ratio and 
extracted the unabsorbed fluxes 
at various epochs in $0.3-2.4$ and $2.0-8.0$ keV ranges. 
We also did spectral analysis of  $\sim 180 ks$  {\it Chandra} ACIS-S 
data in 2005 given its large exposure. 
We get an excellent fit to the data using a two-component thermal
plasma model (reduced $\chi^2$ of 1.09) 
with absorption column density of $(6.0 \pm 1.5) 
\times 10^{20} \, \rm cm^{-2}$.
 This is close to the 
Galactic absorption in the direction of M81, which is expected at such late
epochs when the cool shell absorption may have become insignificant.
The two temperatures obtained from the best fit are
$ 0.73 \pm 0.04$ keV and $2.21 \pm0.24$ keV.
Other models do not fit  the data well.
The Non-equilibrium ionization (NEI)
 model gives  reduced $\chi^2 = 1.88$ for
 104 degrees of freedom and power-law model gives  reduced 
$\chi^2 = 3.48$ for 105 degrees of freedom.
The two temperature bremsstrahlung model yields a 
reduced $\chi^2 = 1.90$ for 103 degrees of freedom.
Hence, the two-component thermal plasma model is the
most plausible one. 
We could not fit the various lines to the data given its sparseness.
We plot the spectral fit to this data in Fig. \ref{fig:spec}.

We also extracted the archival data observed with {\it Chandra}
ACIS-S in HETG
mode at 15 occasions from Feb 24, 2005 to Aug 12, 2006 (Obs IDs:
6174, 6346-47, 5600-01, 6892-6901; PI: Canizares) and analyzed it. These observations
were taken for M81, the host galaxy of SN 1993J which is 3' away from the
SN. SN 1993J flux was a byproduct of these observations. 
Half of the observations were centered on ACIS chip S2 and the
rest of the half on S3. We combined all the datasets in these two major
sets and extracted the flux using best fit models.
In this case, we fixed the column density to be
 $0.06 \times 10^{22}\, \rm cm^{-1}$, the one obtained from the
2005 {\it Chandra} observations (see above).

Latest observations of SN 1993J 
with {\it Chandra} was taken on 2008 Feb 01 (PI: Immler). 
The observations
were taken in ACIS-S mode without any grating. The total 
exposure time was 14.80 ks, out of which we could extract 10 ks of
good data. The count rate in this 
observation was $(2.22 \pm 0.15) \times 10^{-2}$
cts in full ACIS band. We converted these count rate into flux 
using $N_H=0.06 \times 10^{22}\, \rm cm^{-1}$ and temperature of 0.7 keV.

\subsection{XMM-Newton observations}

XMM-Newton observed SN 1993J on 22nd April 2001 around eight years after
the explosion for about 132 ks duration, under 
obsID 0111800301 \citep{zim03}.  Data from the EPIC PN camera, run 
in small window mode, and the MOS2 camera in imaging mode were
obtained. 
\citet{zim03} did the detailed analysis of the data and fit the X-ray spectrum 
with 2-component thermal plasma model. They claimed emissions from highly 
ionized Mg, Si, S, Ar, Ca and complex Fe. We used their best fit parameters to
extract the unabsorbed luminosity in 0.3--2.4~keV, 2--8~keV and 0.3--8~keV
bands.

\subsection{{\it Swift} observations}

The X-Ray Telescope \citep[XRT][]{bur05} on board the {\it Swift}
Observatory \citep{geh04} observed SN~1993J on 2005 April 21,
2005 August 25, 2006 June 24, 2006
November $18-20$ and on 2008 January 09--15. 
We combine the data closely spaced in time, 
i.e. 2005 April 21--August 25, 2006 June 24--November 18, 
and 2008 January 09--15, to increase the 
sensitivity of the data. 

The HEASOFT\footnote{http://heasarc.gsfc.nasa.gov/docs/software/lheasoft/}
(v6.2) and {\it Swift} software (v2.6.1, build 20) tools and latest
calibration
products were used to reanalyze the data.
X-ray counts were extracted from a circular region with a 10~pixel ($24''$)
radius centered on the optical position of the SN. The background was
extracted
locally from a source-free region of $30''$ radius and corrected for the 100\%
 encircled energy radius, to account for the detector and
the sky background, and for residual diffuse emission from the host galaxy.
SN~1993J is detected in X-rays in the 21~ks XRT observation from 2006
November $18-20$
at a $7.0\sigma$ significance of source detection
and a net
count rate of $(4.1\pm0.6) \times 10^{-3}~{\rm cts~s}^{-1}$ (0.2--10~keV).
Table \ref{tab:1} gives details of these observations.

\section{Results and Interpretation}
\label{sec:results}

\begin{figure}
\includegraphics[angle=0,width=0.5\textwidth]{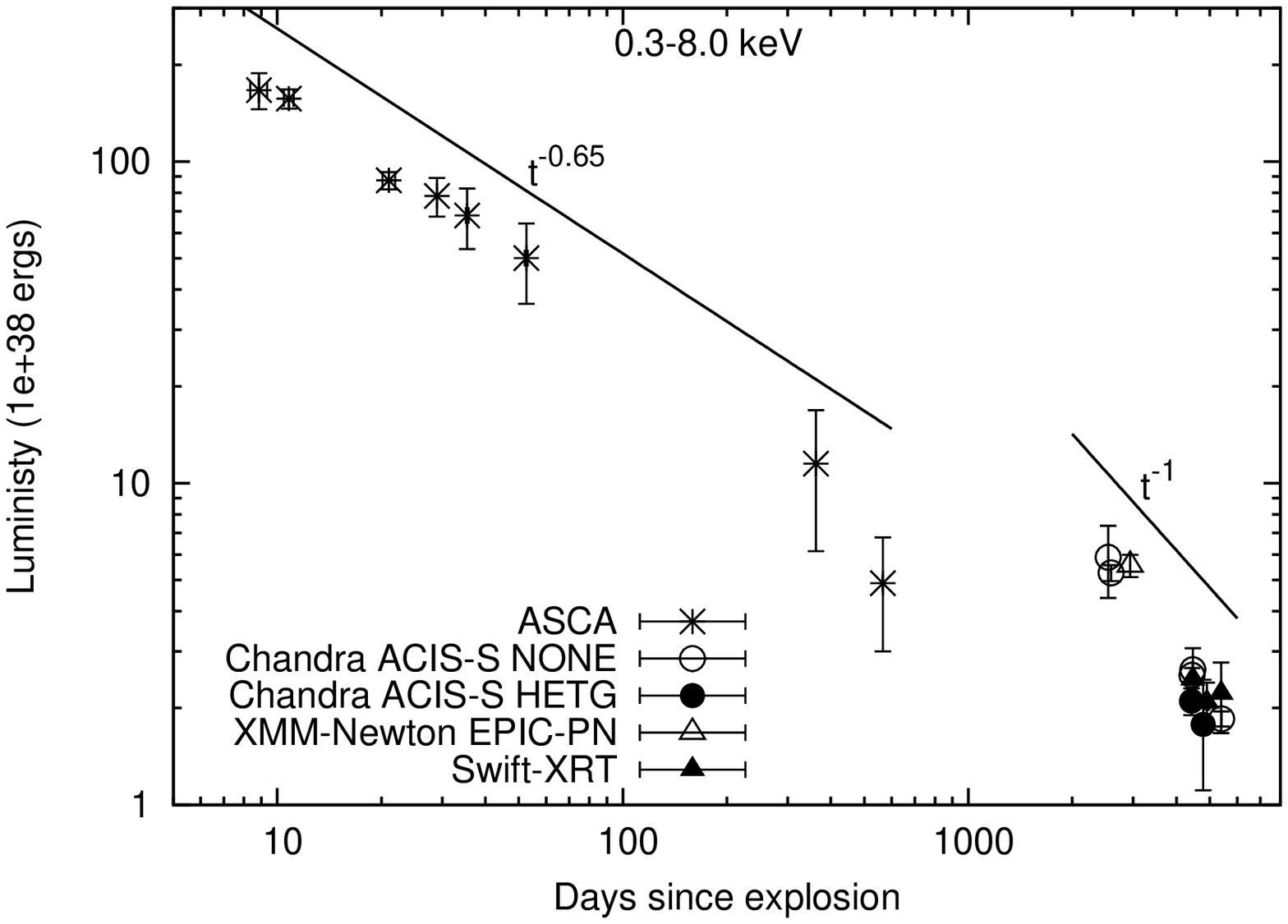}
\includegraphics[angle=0,width=0.5\textwidth]{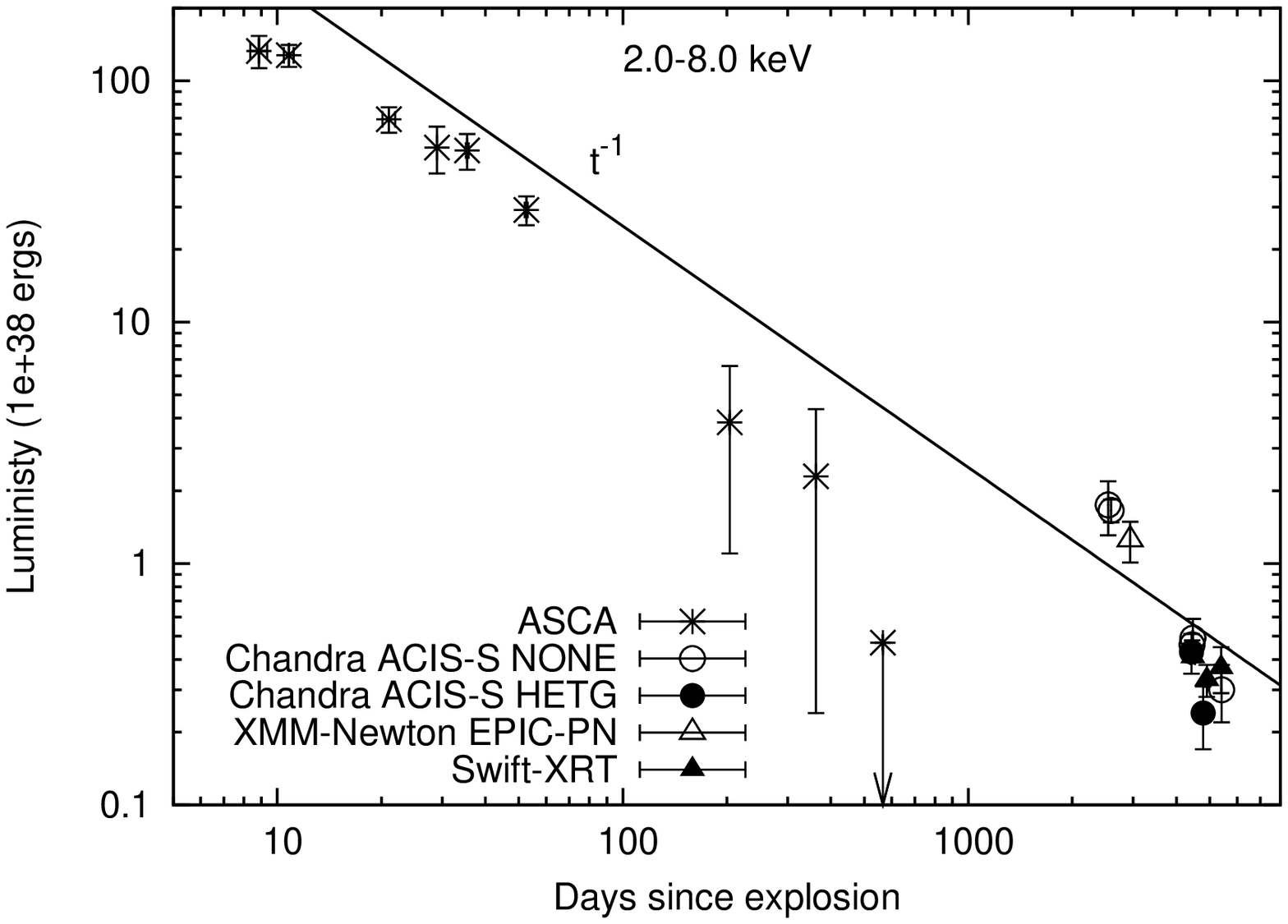}
\includegraphics[angle=0,width=0.5\textwidth]{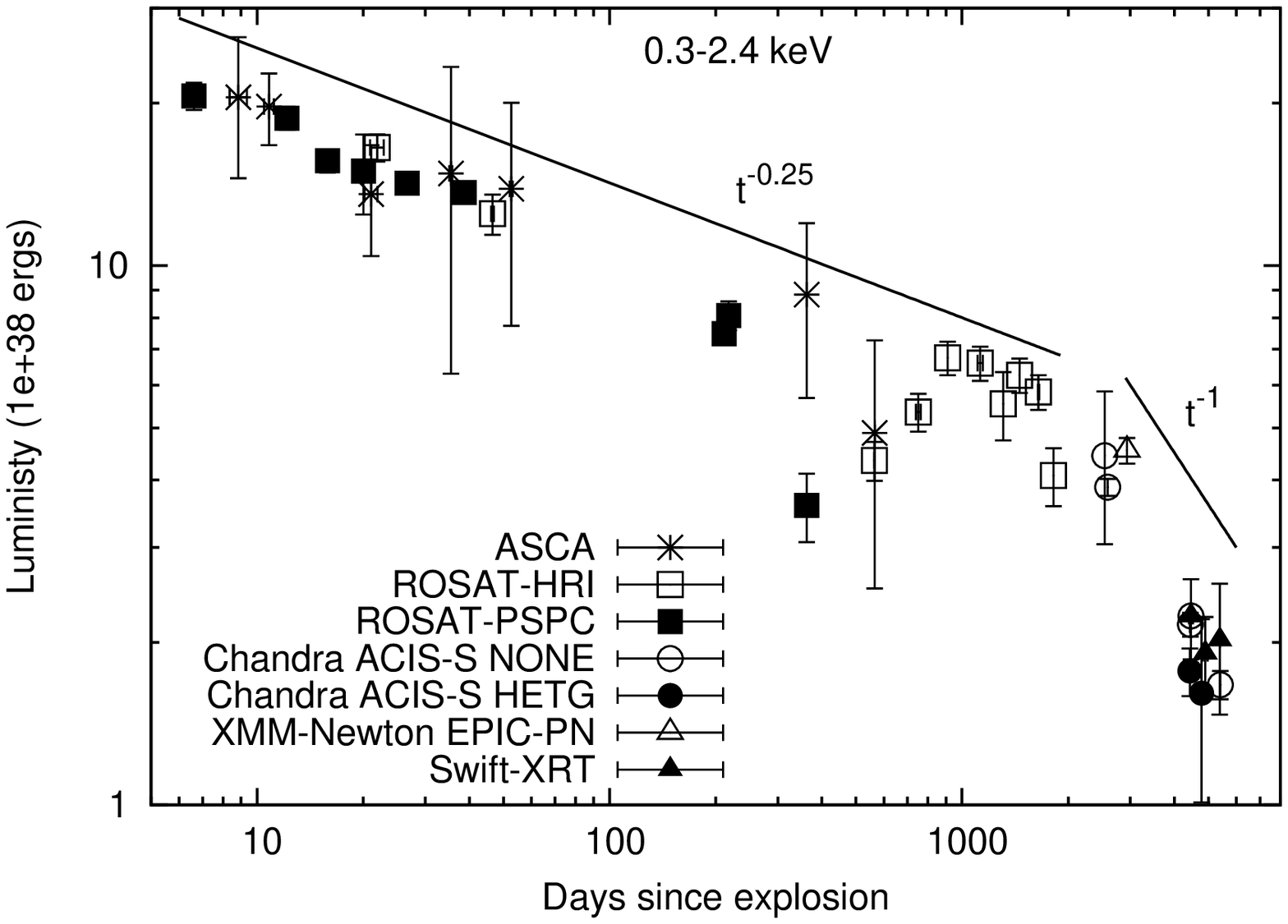}
\includegraphics[angle=0,width=0.47\textwidth]{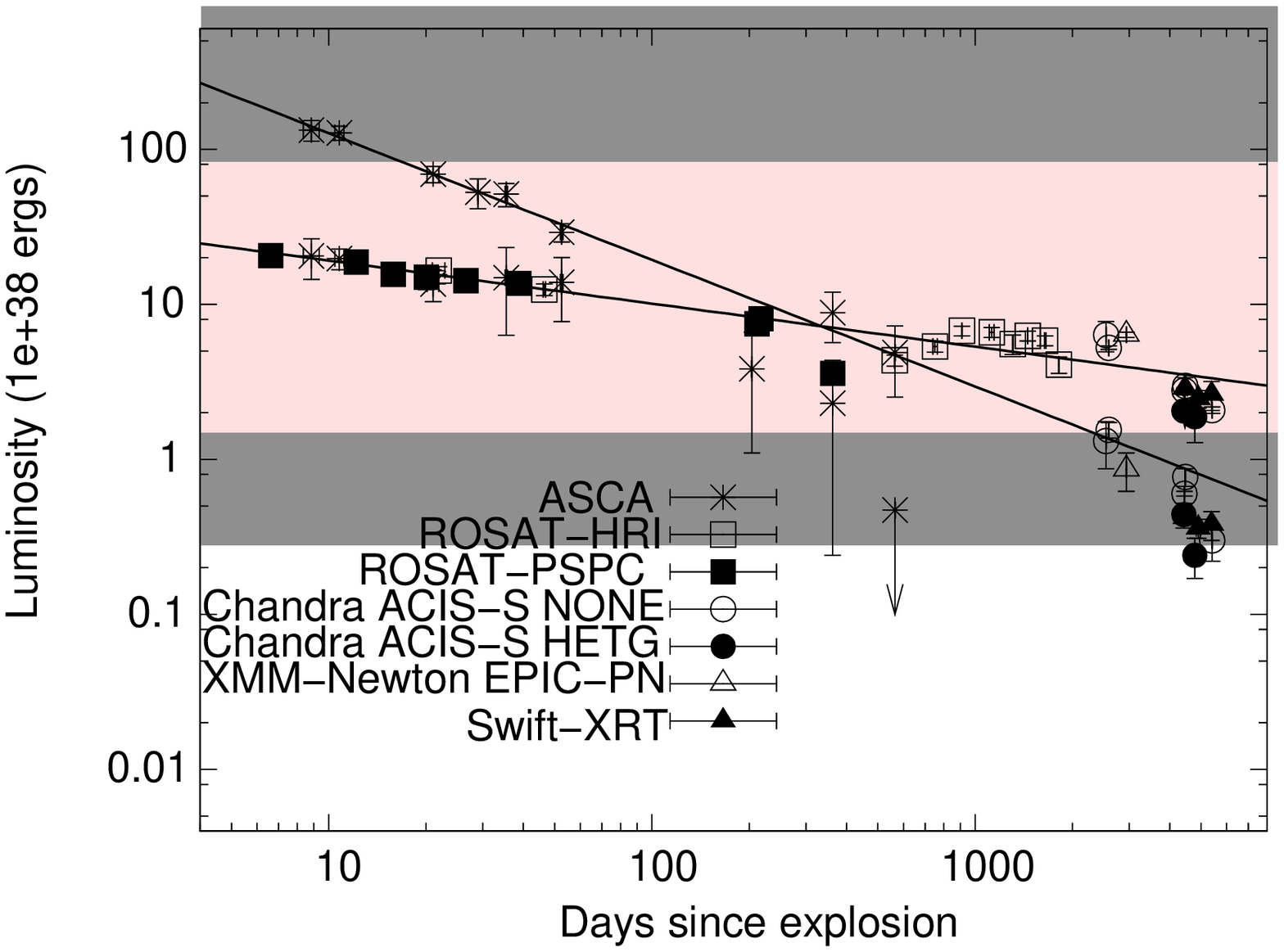}
\caption{ The X-ray light curves of SN 1993J
in the 0.3--8~keV (upper left panel), 2--8~keV  (upper right
panel), and 0.3--2.4~keV (lower left panel) bands
observed with multiple telescopes.
For around first 1000 days, 0.3--2.4~keV light
curves declines slowly ($t^{-0.25}$) whereas the 2--8~keV
light curves decline as $t^{-1}$. The overall 0.3--8~keV light curve declines as 
$t^{-0.65}$ indicating radiative nature of the reverse shock. 
After around day 1000, the shocks seems to become adiabatic.
The lower right panel shows the comparison between the 0.3--2.4~keV and
2--8~keV components and demonstrates that the soft component take overs 
around day 200 and dominates at late epochs.
\label{fig:lc}}
\end{figure}

We tabulate the details of all the above observations 
described in \S \ref{sec:observations} in
Table \ref{tab:obs}. 
We report the unabsorbed luminosities in 
0.3--2.4, 2--8 and 0.3--8~keV bands in Table \ref{tab:1}. 
We plot the light curves 
constructed in these bands from the above observations in Fig. \ref{fig:lc}.
The light curve in 0.3--8~keV 
shows linear decline with powerlaw index of $-0.65$ for first few hundred days.
At late epochs, from day 2500 onwards, the light curve declines at
$t^{-1}$. Due to the lack of observations between day 500 to 2500, 
it is not possible to predict its evolution in this time range. 
However, the flux seems to have increased in this gap of 
$\sim$ 2000 days, contrary  to the expected decline.
The light curve in 2--8~keV seems to have declined much 
faster with a powerlaw index of $-1$ throughout the
all the observations. In this band too, there seems to be a possible 
increase in the luminosity in the gap of 500--2500 days.
Fortunately, 
ROSAT-PSPC observations have covered this gap with many data points 
in the softer band. 
The light curve in 0.3--2.4~keV shows an overall decline in the light curve
with a powerlaw index of $-0.25$ for until about 1500 days.  Thereafter, the 
light curve is consistent with $t^{-1}$ decline. 
However, after day $\sim 200-300$ the light curve reveals a sudden drop in the 
luminosity which rises back slowly and reaches back to the $-0.25$ index 
profile by day 800. 
There is another possible small dip around day 1000-1500.

As discussed in \S \ref{sec:previous}, 
the X-rays may originate either from the forward shock 
or from the reverse shock.
Since the density and 
temperature conditions of the two shocks are widely different, 
this information can help us
identify the X-ray emission regions in a given energy band. 
The CS temperature can be written as
$$T_{CS}=1.36 \times 10^9 \left
(\frac{n-3}{n-2}\right)^2
\left(\frac{V_{ej}}{10^4\, \rm km\,s^{-1}}\right)^2\, \rm K.$$
Radio VLBI of SN 1993J suggests an expansion velocity of the
order of $7000-8000$ km s$^{-1}$ 
at current epoch \citep{bar07,bar02}. H-$\alpha$ observations
suggest that the 
ejecta velocity has to be at least 7500 km s$^{-1}$. \citet{fra96}
have argued that the value of density power law index $n$ is $ 8-12$.
 Putting these values 
in the above equation implies that
CS temperature should be at least $T_{CS}\ge50 $ keV. 
At such high temperature of the CS shock, there will be very little emission
which will come out  in the
0.3--8~keV band.
At early epochs
the velocities 
would be much higher
resulting in high  $T_{CS}$, of the order of at least few tens of keV and thus will 
have little flux in the energies below 8 keV.
\citet{fra96}
have also indicated that all the flux below
10 keV is mostly arising from the reverse shock on the basis of early
X-ray observations of the SN. The above discussion  indicates that the  
 0.3--2.4~keV and 2--8~keV bands are at much lower 
temperatures to account for
the emission from the forward shock;
and most probably the X-rays in these bands
are arising  mostly from the reverse shock. Our claim is further strengthened by
the hydrodynamic simulation (see \$ \ref{sec:hydro}), which also seems to suggest
the reverse shock origin.

X-ray light curve behaviour can give information about the nature
  of the reverse shock from which majority of X-rays is coming.  The
  adiabatic shock luminosity declines with a powerlaw of index of
  $-1$, whereas the radiative reverse shock declines much slowly.  The
  slow decline of 0.3--8~keV light curve at early epoch in consistent
  with the radiative nature of the reverse shock at these epochs.
The late time $-1$ decline is
  consistent with an adiabatic reverse shock. Due to the lack of data
  it is not possible to pin point the transition from radiative shock
  to adiabatic shock.  However the soft band light curve without this
  gap of 2000 days clearly indicates that the shock made the
  transition from a radiative reverse shock to an adiabatic on
  around day 1500--2000.  
Initially hard band emission is much higher than the soft band emission
but the latter
  dominates from day 200 onwards.
Since SN 1993J emission is becoming softer with time,
  this kind of luminosity evolution is expected. The fast decline of
  luminosity in 2--8~keV band also indicates that very little of this
  radiation is being absorbed by the cool shell in this band and most of it is
  coming out as adiabatic shock.  The overall light curve evolution is
  consistent with an early radiative reverse shock turning into an
  adiabatic shock after around 1000 days. This kind of behaviour has
  been predicted by \citet{nym06} for such supernovae.  The sudden
  decline and gradual rise of the X-ray flux between day 200--800 and
  a possible dip around day 1200--1500 may be due to the density
  fluctuations in the ejecta.  \citet{suz95} has predicted a
    possibility of a jump in the SN 1993J light curve. They claim that
    while moving backwards into the ejecta, the reverse shock
    encounters a density jump at the interface of the hydrogen
    envelope and the Helium core.  This density jump at the interface
    of the two regions will show up as luminosity jump in the light
    curve. \citet{mio01} clearly show the effect of this
    density jump on the radio emission. 
 \citet{suz95} have calculated
    the possible time for this jump to be around 500-1000 days,
    although \citet{mio01} find it to be slightly later.
It is probable that we are seeing this effect  in the
 light curves of SN 1993J. However, \citet{suz95} model is 1-D and
the strong density gradients are likely to be smoothed by instabilities.

We  use the cooling time expression in Eq. \ref{eq:tcool} to
derive a timescale for which a reverse shock may remain radiative
for the relevant parameters for SN 1993J mentioned in previous sections.
A rapid cooling of the gas ahead of the shock is required such that at any stage
the cooling timescale is less than the elapsed timescale so that a
layer of cool and dense absorbing gas forms, i.e.  $ t_{\rm cool} / t
< 1$.  This condition along with Eq. \ref{eq:tcool} gives
\begin{equation}
t   \le   \frac{ (n-3) (n-4) (n-2)^{3.34} }{605} 
\left( \dot M_{-5} \over u_{\rm w1} \right)
 \left( V_{ej} \over 10^4 \, \rm km \, s^{-1}\right)^{-5.34}\, \rm days.
\label{eq:cool}
\end{equation}
For $n=12$, this equation gives 
$t\le 260(\dot M_{-5}/u_{\rm w1})(V_{ej}/10^4 \, \rm km \, s^{-1})^{-5.34}$ 
days. We tabulate radiative timescales for various values of
$V_{ej}$ and $\dot M_{-5}$ in Table \ref{tab:tcool}. 
This indicates that it is
possible for reverse shock to remain radiative at  late epochs,
as indicated by our light curves.

\section{Hydrodynamic simulation and computed X-ray light curves}
\label{sec:hydro}

The analytic calculations in this paper assume a CS density profile that
varies as $r^{-2}$.
However, in numerical simulations, we do not constrain ourselves with
$r^{-2}$ and explore a larger parameter space. We have carried out
several analytic simulations to find the best possible CSM density
structure that provides an adequate fit to the observed X-ray data.
 We chose to reproduce the data in the hard X-ray band,
  because this band is likely to behave adiabatically (as evident from
  our 2--8~keV light curve) and hence have no complicated line
  emission due to the cooling effects. Our simulations are based on
  \citet{mio01} for a spherically symmetric system using high
  resolution VH-1 three dimensional finite difference code.
\citet{mio01} produced simulated radio light
  curves using a detailed radiative transfer calculation and fit the
  radio light curve of SN 1993J.  These hydrodynamic results were
  updated in \citet{bar07} to match updated VLBI data from
  1993J. Herein we use an updated version of this hydrodynamic
  calculation to simulate the observed X-ray light curves of SN 1993J
  reported in this paper.  The code computes the interaction of the
  ejecta with a CSM whose density profile
leads to the calculation of X-ray flux
  density of the SN at each epoch.  The calculations were carried out
  in one dimensional Lagrangian coordinates and remapped on to an
  Eulerian grid.  The shocked interaction region between the forward
  and reverse shocks was adequately resolved, which is important, as
  most of the X-rays are emitted in this region.  

The hydrodynamic runs use the 4H47 ejecta density distribution model
of Shigeyama and Nomoto (courtesy of K. Nomoto 1999, private
communication). This model had an ejecta mass of $3.12 \; M_{\odot}$,
the pre-SN progenitor star radius of $350 \; R_{\odot}$ and an
envelope mass of $0.47 M_{\odot}$. The mass fraction of helium in the
envelope was about 0.79.  The explosion kinetic energy was $10^{51}$
erg. The surrounding density profile has a slope that varies over
time, from r$^{-1.4}$ in the very early stages, to r$^{-2.1}$ up to
10$^{17}$ cm, and then a steeper drop to r$^{-2.6}$ at late epochs.
This CSM density profile was used to adequately fit the VLBI data
  in \citet{bar07}, and produced radii and velocities comparable to
  the radio and optical (H$\alpha$) observations. 
The computation of radiative emission from the hydrodynamic model
further requires the electron temperature, whereas the the
hydrodynamics gives the post shock temperature, which is the
temperature of the ions $T_{ion}$.  We have found that an electron
temperature about 15 \% of the ion temperature provides an adequate
fit to the emission.  This is a somewhat simplistic assumption.
We calculated the X-ray emission between 1--6~$\AA$ (2--10~keV).
The observational analysis show that the
absorption by the cool shell is less significant
for the hard X-ray emission, thus we ignore it. Line emission may also
be ignored in this hard band.
We used abundances that are appropriate
for the ambient medium and SN ejecta in SN 1987A, as outlined by
\citet{lun99}.  \citet{nym09} also found the
SN 1987A abundances to be a reasonable set of abundances
for the SN 1993J. The ionization of various species were
computed by a model due to \citet{shu82}.  
Differences between various ionization models are small 
and do not alter the hard X-ray flux significantly, or the fundamental result
that the X-ray flux arises from reverse shocked ejecta.

\begin{figure}
\begin{center}
\includegraphics[angle=90,width=0.9\textwidth]{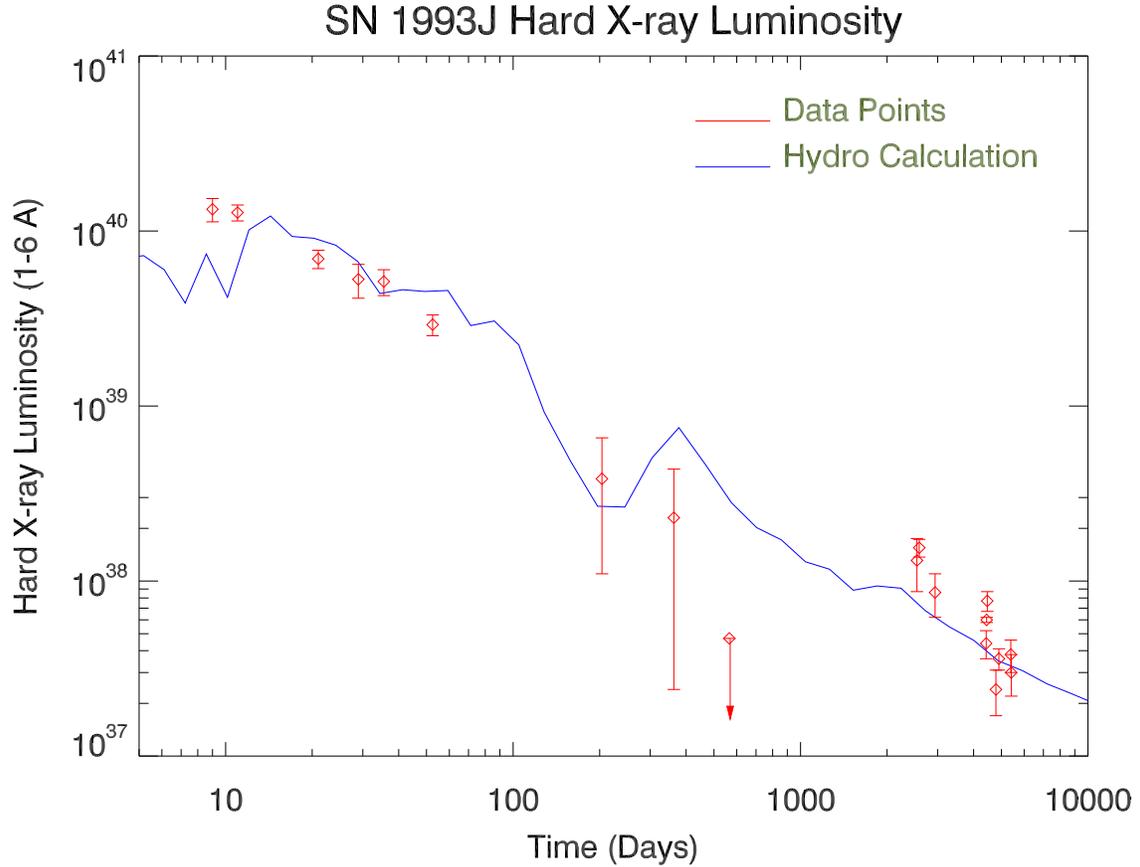}
\caption{Model plot of the
Hard X-ray band (1-6 $\AA$)
 luminosity evolution calculated
from hydrodynamic modelling with VH-1 and CHIANTI codes starting from a Nomoto and
Shigeyama 4H47 model. We plot this model with the hard band
observed luminosities in 2--8~keV band.
The model does not trace the early luminosity
evolution for first 10--15 days (probably due to
complicated X-ray emission process at early phase),
however
seems to reproduce the
observed light curve well after that.}
\label{fig:hydroresults}
\end{center}
\end{figure}

Given the density and temperature in every
zone, the hydrodynamic model calculates the free-free and bound-free luminosity from
that zone using the CHIANTI code \citep{chi97}.  This is being done
for 10000 zones, over each time-step. The contribution from each zone
is added up to give the total luminosity. Having the contribution from
each zone also allows us to determine exactly the part of the density
profile that is contributing most to the X-ray emission. 
The computed X-ray
luminosity in the hard band is shown in Fig.
\ref{fig:hydroresults}. Looking at the simplicity of our run, the
match between the simulated light curve and actual luminosity is encouraging. 
Between 10 and about 400 days the luminosity
decreases only slowly.  Some of the later peaks are due to the reverse
shock running into ejecta structures which compresses them.
After 1000 days, the slope becomes steeper ($L \propto
t^{-1}$), in accordance with the data.
We find that the hard X-rays arise from just
behind the reverse shock. The region of hard X-ray production expands
inward in the ejecta over the next decade or so till they begin to
arise from most of the shocked ejecta, between the reverse shock and
the contact discontinuity. After about 50 years, the forward shock
starts to dominate the emission.
This confirms
our prediction that almost all the X-rays below 8~keV over the
observed time period are coming from
the reverse shock.  

Our hydrodynamic simulation models are carried out in spherical
symmetry.  It is encouraging that even using these simple
hydrodynamic models it is possible to understand qualitatively the
evolution of the SN, and explore several basic features, such as the
fact that the hard X-ray emission is coming predominantly from the
reverse shock.

\section{Discussion}
\label{sec:discussion}

\subsection{Column Depth and the cool shell}
\label{sec:column-depth}

\citet{imm01} had analyzed ROSAT data fixing the
column depth to be $4\times 10^{20}$ cm$^{-2}$, same as the
Galactic absorption column density. However, the reanalysis
of the same data by \citet{zim03} and letting column density to be a 
free parameter revealed that column densities are much higher than the Galactic
column density. 
\citet{uno02} fit the
ASCA data with two component thermal plasma model and found that the
low temperature component has much higher column density than the high temperature
component. 
These high column densities in excess to the
Galactic absorption can be attributed to the absorption by an
additional cool shell.
Since most of the 
X-ray emission below 8~~keV is coming from the 
reverse shock, 
this indicates that both the ROSAT and the ASCA data have revealed the 
presence  of a
cool shell between the forward and a reverse shock, 
thus presence of a radiative reverse shock. 
However, the column density seems to reduce with time and after around 2000 days, 
the best fit column densities are consistent with that of
the Galactic absorption. This indicates that the
reverse shock has most likely become adiabatic by this epoch.

\subsection{Hardness Ratio and electron temperature}
\label{sec:hardness-ratio-Te}

\begin{figure}
\begin{center}
\includegraphics[angle=0,width=0.9\textwidth]{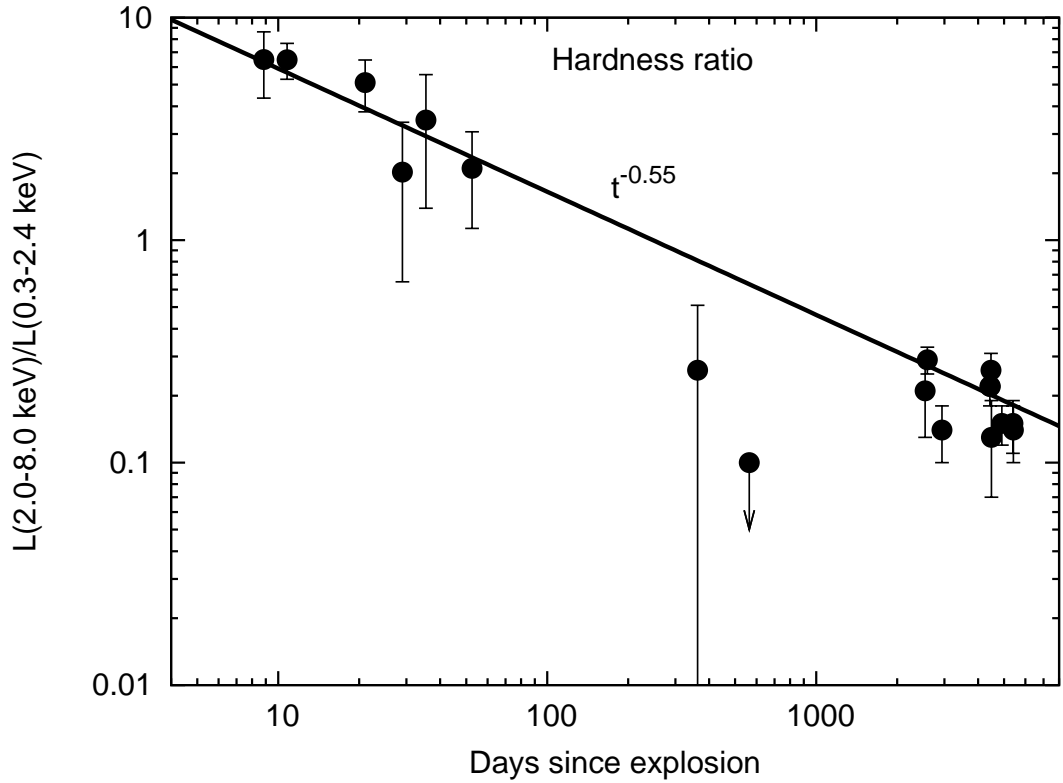}
\caption{Here we plot the ratio of 2--8~keV luminosities versus
0.3--2.4~keV luminosities at various epochs.
The hardness ratio decreases with time and
roughly follows a powerlaw dependence with time index of $-0.55$.}
\label{fig:hardness}
\end{center}
\end{figure}

In Fig. \ref{fig:hardness}, we plot the ratio of 
SN~1993J luminosity in the 2--8~keV  versus 0.3-2.4~keV bands.
The figure clearly shows that the hard X-ray emission
dominates for first $\sim 100-200$ days. After $\sim 200$ days, 
soft X-ray emission
starts to take over and continues to dominate. At current epoch, i.e.
15 years after
the explosion, the soft X-ray is dominant by around an 
order of magnitude.

Since most of the flux below 8 keV is coming from the 
reverse shock, the 
ratio of the reverse shock free-free luminosity in  
2--8~keV band to 0.3--2.4~keV band can be written as 
\citep{fra96}:
\begin{equation}
\mathcal H \equiv \displaystyle\frac{L_{2.0-8.0 \rm keV}}{L_{0.3-2.4 \rm keV}}=
\displaystyle\frac{\int_{2.0 \rm keV}^{8.0 \rm keV}  (\frac{E}{kT})^{-0.4}
\exp(-\frac{E}{kT}) {\rm d} E}
{\int_{0.3 \rm keV}^{2.4 \rm keV} (\frac{E}{kT})^{-0.4}
\exp(-\frac{E}{kT}) {\rm d} E}
\label{Eq:hardness}
\end{equation}
Here $(E/kT)^{-0.4}$ is the Gaunt factor  in the given
energy regime. This equation puts some very interesting
constraint on the temperature of the reverse shock.
The current hardness ratio of 0.14 
corresponds to the electron temperature of 1.05 keV. 
This equation also shows that electron temperature was $\sim 1.5$ keV
around day 500 and it has evolved very slowly since then to
reach the current value of 1 keV. Since free-free emission dominates
above temperature 2 keV \citep{fra96}, our derived temperature 
of 1 keV at current epoch will contribute 
mostly to the free-free continuum emission, as well as 
possibly to the line emission. 

For maximum electron temperature of $\sim 30$ keV \citep{fra96}, 
the hardness ratio is
$1.45$. In fact,  
the Eq. \ref{Eq:hardness} above reaches asymptotic
 upper limit of $\sim 1.5$ which is
around day 100, irrespective
of temperature. However, this demonstrates
 that in the initial X-ray observations, where
hardness ratio is larger than 1.5, the contribution between $2-8$ keV is not
solely because of the reverse shock but may have some contribution from the
CS shock too. 
After day $\sim 100$, 
all the X-ray flux below 8 keV
comes from the reverse shock (see Figure \ref{fig:hardness}).

\subsection{H$\alpha$ and the X-ray luminosity evolution}
\label{sec:Halpha-evolution}

H$\alpha$ emission in SNe arises 
initially by radioactivity and later  by reprocessing of X-rays
produced due to ejecta-wind interaction. In case of
adiabatic shocks, the H$\alpha$ emission may come from the unshocked material
heated by the X-rays coming from the adiabatic shock waves.
If the reverse shock is radiative, then the dense cool shell between the
reverse and the forward shock may also give rise to significant 
H$\alpha$ emission.

There have been extensive H$\alpha$ observations of
SN 1993J. 
\citet{mat00,mat00b} describe the detailed optical spectra of SN 1993J
up to 2500 days. 
In \citet{mat00b}, they also measure the 
linewidths of boxy H$\alpha$ profile,
an indication of the CS
 interaction,  from day 433 onwards.
\citet{pat95} reports H$\alpha$ fluxes between 
day 171 to day 367.
\citet{zha04} report photometric observations of SN 1993J
from the year 1995 to 2003. 
\citet{hou96} showed that the decay of $^{56}Ni$
provides the ultimate power source for H$\alpha$ emission up to 250
days.  After 250 days, continued presence of strong broad
lines of H$\alpha$ in SN 1993J have been explained as a result of
photoionization by X-rays and UV emission from the radiative reverse
shock propagating into the supernova ejecta \citep{fra05}.
\citet{pat95}
show that late time H$\alpha$ emission from SN 1993J can be
described by a shell of hydrogen between 7500 to 11,400 km s$^{-1}$.
Similar velocity range for emitting gas shells for broad, box-shaped
UV lines are seen in the HST spectra of SN 1993J \citep{fra05} 
apparently coming from an ejecta and a cool dense shell.

\begin{figure}
\begin{center}
\includegraphics[angle=0,width=0.9\textwidth]{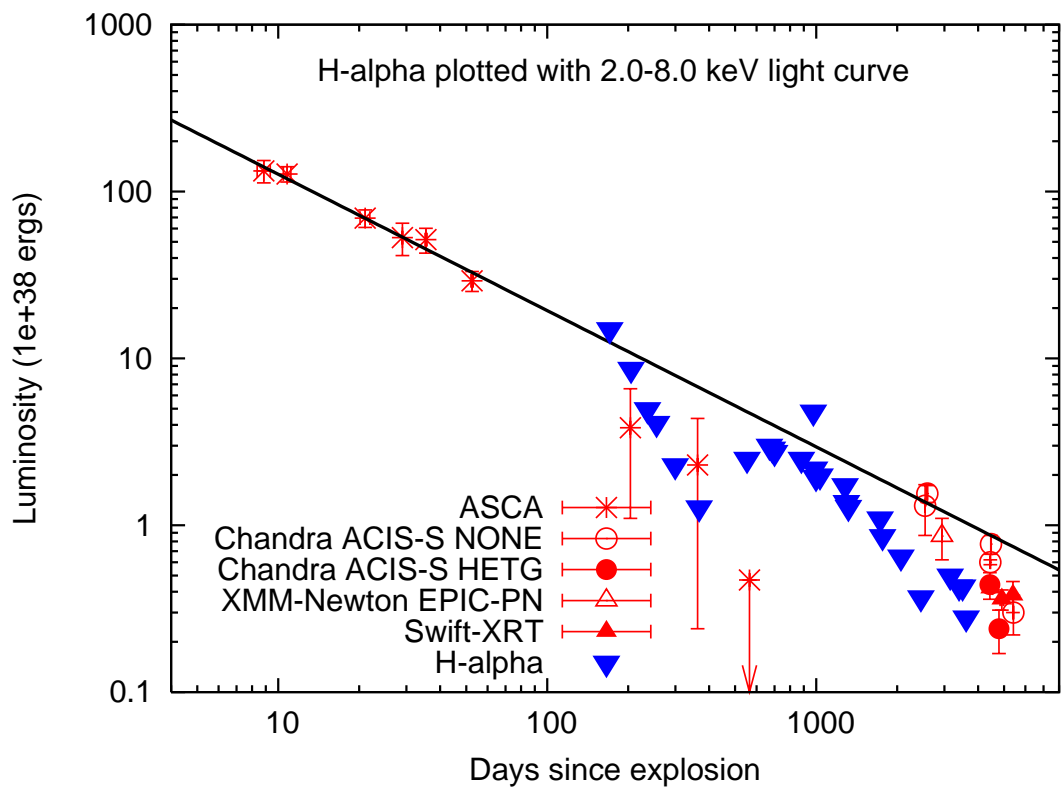}
\caption{ Plot of H$\alpha$ luminosity
light curve (taken from Table \ref{tab:halpha}). We also plot the
2--8~keV X-ray light curve. The H$\alpha$
roughly traces the 2--8~keV band light curve
after day $\sim 300$ indicating that  H$\alpha$ arises due to
the CS interaction. The H$\alpha$ luminosities are high
fractions (30--50\%) of X-ray luminosities, which probably indicate
towards the presence of clumps in the ejecta.}
\label{fig:halpha}
\end{center}
\end{figure}

We tabulate
these H$\alpha$ luminosities
after day 170  
in Table \ref{tab:halpha}. We also plot the H$\alpha$ luminosities 
with soft and hard band X-ray luminosities in figure \ref{fig:halpha}.
The figure shows that the 
 H$\alpha$ luminosity seems to follow the  
 hard X-ray band luminosity evolution after a few hundred days, but 
with significant efficiency. However,
it fails to trace the soft band X-rays completely.
The significant 
rate of conversion of the kinetic luminosity
of the ejecta wind interaction into broad H$\alpha$ emission by the reprocessing
of the X-ray luminosity of the reverse shock wave was 
also
noted by \citet{pat95}. 
It should be noted that \citet{pat95} have cautioned that possibly up to
30\% of the emission at H$\alpha$ wavelengths in SN 1993J may originate from an
unidentified band of emission around 6600 $\AA$ which is hypothesized due to a
broad blend of emission lines extending between 6050-6800 $\AA$ as seen to be
present in type Ib/Ic SNe.

The H$\alpha$ seems to trace the hard band evolution but the efficiencies are
30-50\% of the X-ray production in this band, whereas the expected
efficiency is typically no  more than 1-5\% of the X-ray luminosity. This indicates that
there could be another component which is contributing significantly to H$\alpha$
flux. The clumps in the ejecta can be one such candidate which can give rise to
H$\alpha$ with high efficiency. The non-smooth H$\alpha$ evolution also
probably indicates towards the possibility of significant H$\alpha$ origin
from the clumps. 
\citet{pat95} also invoked an earlier suggestion by
\citet{chu93} where the clumpiness of the
wind material and possibly the clumpy structure of the dense and Rayleigh 
Taylor unstable region in the ejecta helped attain a more efficient 
transformation of kinetic energy into radiation. 
\citet{chu94} proposed a model for SN 1988Z, in which 
a radiative shock wave is driven into
the dense cloud by thermal and dynamical pressures behind the main 
(blast wave shock and
reverse shock) 
shock waves or by the dynamical pressures of the expanding unshocked ejecta.
The shock in the cloud is significantly slower due to the higher density in the
cloud and the shocked gas cools by soft X-ray or UV emission which then pumps
the optical emission from the cool dense material behind the radiative
shock wave in the cloud.

\section{Conclusions}
\label{sec:conclusion}

In this paper, we have presented the complete
X-ray light curves of SN 1993J in 
0.3--2.4~keV, 2--8~keV, and 0.3--8~keV bands.
We demonstrate that most of the emission below 8~keV comes from the
reverse shock, except for very early soft band emission when the cool shell
absorbed all the soft flux from the reverse shock. The  light curves
reveal that the
 reverse shock is radiative for around initial 1000 days and then it becomes 
adiabatic at later epochs, as expected for such supernovae \citep{nym06}. 
The evolution of column density too seem to indicate this. The column density was
 higher than the Galactic column density during the ROSAT, ASCA and early Chandra
and XMM observations, and later became comparable to the Galactic absorption,
indicating presence of a cool shell during around first 1000 days. We
demonstrate that for SN 1993J parameters, it is possible for the
shock to be radiative at such late epochs.

We have
  carried out numerical hydrodynamic computations, and calculated the
  hard X-ray flux from the same, which agree reasonably well with the
  observed data.
Our simulations clearly show that all the emission below 8 keV is indeed
coming from the reverse shock, strengthening our claim.

The large fraction of H$\alpha$ flux in comparison to the X-ray emission
indicates possibility of clumps in the ejecta. 

One of the important questions is how these young supernovae evolve
  towards a Supernova Remnant. 
Since young SNe and older SNRs
  are both products of explosive events a natural question is whether
  the former evolve continuously into the latter with time, whether
  similar emission mechanisms differing only in length-scales
  and time-scales operate or if the SNe fade away only to switch on later
  as SNRs with a different mechanism of radiation.  
One of the very well
  studied SNR in our galaxy, Cassiopeia A was thought to be of the
  Type IIb or IIn \citet{che-oishi03}.  However, it has recently been
  conclusively classified as a Type IIb SN, the same type as that of SN
  1993J \citep{kra08}. 
Late time observations of SN
  1993J may provide an important link between young SNe and SNRs.  
SN 1993J had a large X-ray flux,
  which, coupled with it being a nearby SN, made it easily observable
  for a long time with a well-sampled light curve.  This has made SN
  1993J one of the best studied extragalactic supernovae, and provided
  a wealth of information on SN evolution and their circumstellar
  interaction. Further observations of this SN for a long time is 
sure to provide much valuable information on these questions.

\acknowledgements 

We thank the anonymous referee for useful comments.
P.C. thanks Roger Chevalier for useful discussions.  We thank
Richard McCray for discussions and his comments on an earlier draft of
this manuscript and thank Claes Fransson, Roger Blandford, Alex
Filippenko, Subir Sarkar, Stephen Smartt for several illuminating
discussions. We embarked on this work at the 2007 Aspen Workshop on
``Supernova 1987A: 20 Years After" and we thank the organizers of that
Workshop. A.R. thanks the participants of the Darjeeling School and
Workshop on Supernovae and GRBs in May 2008 where this work was
discussed and Manjari Bagchi and Sayan Chakraborti for their technical
assistance.  P.C. is a Jansky fellow at National Radio Astronomy
Observatory.  The National Radio Astronomy Observatory is a facility
of the National Science Foundation operated under cooperative
agreement by Associated Universities, Inc.  At Tata Institute this
research was supported by the Eleventh Five Year Plan Project
No. 11P-409.  V.V.D. received support from award \# AST-0319261 from
the NSF.  This research has made use of data obtained through the High
Energy Astrophysics Science Archive Research Center On-line Service,
provided by the NASA/Goddard Space Flight Center.  We thank the
Chandra X-Ray Observatory team for carrying out the observations for
Feb 2008 under guest observing program.  CHIANTI is a collaborative
project involving the NRL (USA), RAL (UK), MSSL (UK), the Universities
of Florence (Italy) and Cambridge (UK), and George Mason University
(USA).

\clearpage

\begin{deluxetable}{lllll}
\tablecaption{Details of X-ray observations of SN 1993J
\label{tab:obs}}
\tablewidth{0pt}
\tablehead{
    \colhead{Date of} & \colhead{Mission} &  \colhead{Instrument}&
 \colhead{Observation} &  \colhead{Exposure}\\
    \colhead{observation} & \colhead{} &
 \colhead{} & \colhead{ID} &  \colhead{(ks)}
}
\startdata
1993 Apr 03.41 & ROSAT & PSPC & RP180015N00 & -- \\
1993 Apr 05.25--Apr 06.04 & ASCA & -- & 15000120 & 27.4  \\
1993 Apr 07.25--Apr 07.94 & ASCA & -- & 15000130 & 28.3\\
1993 Apr 08.34--Apr 09.35 & ROSAT & PSPC & RP180015N00 &-- \\
1993 Apr 12.23--Apr 13.01  & ROSAT & PSPC & RP180015N00 &-- \\
1993 Apr 16.83 & ROSAT & PSPC & RP180015N00 & 5.1\\
1993 Apr 16.94--Apr 18.73 & ASCA & -- & 15000030 & 66.4 \\
1993 Apr 17.68--Apr 19.68  & ROSAT & HRI & RH600247A01 & -- \\
1993 Apr 22.05--Apr 24.0 & ROSAT & PSPC & RP180015N00 & --\\
1993 Apr 25.76  & ASCA & -- & 15000020 & 11.1\\
1993 May 01.93--May 02.59   & ASCA & -- & 15000040 & 21.6 \\
1993 May 04.30--May 06.30  & ROSAT & PSPC & RP180015A01 & -- \\
1993 May 12.85--May 13.48  & ROSAT & HRI & RH600247 & --\\
1993 May 18.87--May 19.86 & ASCA & -- & 15000050 & 39.3 \\
1993 Oct 24.60--Oct 25.59 & ASCA & -- & 10018000 & 39.6\\
1993 Nov 01.68 &  ROSAT & PSPC & RP180035N00 & --\\
1993 Nov 07.90 &  ROSAT & PSPC & RP180035A01 & 20.0 \\
1994 Apr 01.45  & ROSAT & PSPC & RP180050N00 & 108.3\\
1994 Apr 01.68  & ASCA  & -- & 51005000 & 32.6 \\
1994 Oct 19.17--Oct 21.31  & ROSAT & HRI & RH600739N00 & --\\
1994 Oct 21.18--Oct 22.33 & ASCA & -- & 52009000 & 46.4\\
1995 Apr 13.71--May 04.90 &  ROSAT & HRI & RH600740N00 & --\\
1995 Oct 19.00 & ROSAT & HRI & RH600881N00 &  -- \\
1996 Apr 15.88--May 07.83  & ROSAT & HRI & RH600882N00 & --\\
1996 Nov 04.50 &  ROSAT & HRI & RH600882A01 & --\\
1997 Mar 31.10  & ROSAT & HRI & RH601001N00 &  --\\
1997 Sep 30.44--Oct 16.00&  ROSAT & HRI & RH601002N00 &-- \\
1998 Mar 26.10 & ROSAT & HRI & RH601095N00 &  -- \\
2000 Mar 21.03 & {\it Chandra} & ACIS-S (NONE) & 390 & 2.4 \\
2000 May 07.23 & {\it Chandra} & ACIS-S (NONE) & 735 & 50.6 \\
2001 Apr 22.32 & XMM-Newton  & EPIC-PN  & 0111800301 & 132.0 \\
2005 Aug 14.41 & {\it Chandra} & ACIS-S (HETG) & 5600 & 36.4\\
2005 Jul 19.60 & {\it Chandra} & ACIS-S (HETG) & 5601 & 84.1\\
2005 Jul 14.81 & {\it Chandra} & ACIS-S (HETG) & 6347 & 64.7\\
2005 Jul 14.07 & {\it Chandra} & ACIS-S (HETG) & 6346 & 55.2\\
2005 Feb 24.29 & {\it Chandra} & ACIS-S (HETG) & 6174 & 45.2 \\
2006 Jul 13.57 & {\it Chandra} & ACIS-S (HETG) & 6899 & 15.1 \\
2006 Feb 08.85 & {\it Chandra} & ACIS-S (HETG) & 6892 & 15.0 \\
2006 Mar 05.99 & {\it Chandra} & ACIS-S (HETG) & 6893 & 15.0 \\
2006 Apr 01.44 & {\it Chandra} & ACIS-S (HETG) & 6894 & 15.0 \\
2006 May 14.54 & {\it Chandra} & ACIS-S (HETG) & 6896 & 15.0 \\
2006 Jun 09.76 & {\it Chandra} & ACIS-S (HETG) & 6897 & 15.0 \\
2006 Aug 12.68 & {\it Chandra} & ACIS-S (HETG) & 6901 & 15.0 \\
2006 Jun 28.98 & {\it Chandra} & ACIS-S (HETG) & 6898 & 14.9 \\
2006 Apr 24.35 & {\it Chandra} & ACIS-S (HETG) & 6895 & 14.7 \\
2006 Jul 28.46 & {\it Chandra} & ACIS-S (HETG) & 6900 & 14.6 \\
2005 Jul 03.06 & {\it Chandra} & ACIS-S (NONE) & 5948 & 12.2 \\
2005 Jul 06.33 & {\it Chandra} & ACIS-S (NONE) & 5949 & 12.2 \\
2005 Jun 18.48 & {\it Chandra} & ACIS-S (NONE) & 5943 & 12.2 \\
2005 Jun 26.90 & {\it Chandra} & ACIS-S (NONE) & 5946 & 12.2 \\
2005 Jun 01.36 & {\it Chandra} & ACIS-S (NONE) & 5937 & 12.2 \\
2005 Jun 09.29 & {\it Chandra} & ACIS-S (NONE) & 5940 & 12.1 \\
2005 Jun 15.06 & {\it Chandra} & ACIS-S (NONE) & 5942 & 12.1 \\
2005 Jun 03.95 & {\it Chandra} & ACIS-S (NONE) & 5938 & 12.0 \\
2005 Jun 06.64 & {\it Chandra} & ACIS-S (NONE) & 5939 & 12.0 \\
2005 Jun 11.89 & {\it Chandra} & ACIS-S (NONE) & 5941 & 12.0 \\
2005 Jun 21.22 & {\it Chandra} & ACIS-S (NONE) & 5944 & 12.0 \\
2005 Jun 24.25 & {\it Chandra} & ACIS-S (NONE) & 5945 & 11.7 \\
2005 May 28.83 & {\it Chandra} & ACIS-S (NONE) & 5936 & 11.6 \\
2005 May 26.16 & {\it Chandra} & ACIS-S (NONE) & 5935 & 11.1 \\
2005 Jun 29.56 & {\it Chandra} & ACIS-S (NONE) & 5947 & 10.8 \\
2005 Apr 21.00 & {\it Swift} & XRT & 00035059001 & 1.6\\
2005 Aug 25.05 & {\it Swift} & XRT & 00035059002 & 5.1\\
2006 Jun 24.00 & {\it Swift} & XRT & 00035059003 & 4.2\\
2006 Nov 18.04 & {\it Swift} & XRT & 00035059004 & 21.2\\
2008 Jan 09.04 & {\it Swift} & XRT & 00036557001 & 4.4 \\
2008 Jan 11.59 & {\it Swift} & XRT & 00036557002 & 3.6\\
2008 Jan 15.47 & {\it Swift} & XRT & 00036557003 & 1.4 \\
2008 Feb 01.00 & {\it Chandra} & ACIS-S (NONE) & 9122 & 10.0\\
 \enddata
 \end{deluxetable}

\clearpage

\begin{deluxetable}{llllll}
\tabletypesize{\scriptsize}
\tablecaption{Unabsorbed X-ray luminosities of  SN 1993J at various epochs
\label{tab:1}}
\tablewidth{0pt}
\tablehead{
    \colhead{Date of} & \colhead{Days since} & \colhead{instrument}&
    \multicolumn{3}{c}{Luminosity ($10^{38}$ erg)}\\
    \colhead{observation} & \colhead{explosion} & \colhead{} & \colhead{0.3--2.4 keV} & \colhead{2--8 keV} & \colhead{0.3--8 keV}
}
\startdata
1993 Apr 03.41 & 6.61 & ROSAT-PSPC & $20.63 \pm 1.20$ & \ldots &\ldots  \\
1993 Apr 05.25--Apr 06.04 & 8.45--9.24  & ASCA & $20.50 \pm 5.99$  & $133.02 \pm 20.27 $ & $166.85 \pm 21.29 $  \\
1993 Apr 07.25--Apr 07.94 & 10.45--11.14  & ASCA & $19.71 \pm 3.00$ & $127.48 \pm13.31 $& $156.91 \pm 10.72 $ \\
1993 Apr 08.34--Apr 09.35 & 11.54--12.55  & ROSAT-PSPC & $18.76 \pm
0.92$ & \ldots &\ldots  \\
1993 Apr 12.23--Apr 13.01 & 15.43--16.24  & ROSAT-PSPC &
$15.63 \pm 0.76$ & \ldots &\ldots  \\
1993 Apr 16.83 & 20.03 & ROSAT-PSPC & $14.96 \pm 2.52$  & \ldots &\ldots  \\
1993 Apr 16.94--Apr 18.73 & 20.14-21.93  & ASCA & $13.56 \pm 3.15$  & $69.27 \pm 8.32$& $ 87.37 \pm 5.36 $ \\
1993 Apr 17.68--Apr 19.68 & 20.88--22.88  & ROSAT-HRI & $16.54 \pm 0.96$
& \ldots &\ldots  \\
1993 Apr 22.05--Apr 24.0 & 25.25--27.20  & ROSAT-PSPC &
$14.20 \pm 0.63$ & \ldots &\ldots  \\
1993 Apr 25.76 & 28.96 & ASCA & \ldots  & $52.91 \pm 11.62$ & $78.22 \pm 10.72 $ \\
1993 May 01.93--May 02.59 & 35.13--35.79  & ASCA &
$ 14.82 \pm 8.52 $ & $51.43 \pm 8.68$ & $67.97 \pm 14.51 $ \\
1993 May 04.30--May 06.30 & 37.50--39.50  & ROSAT-PSPC &
$13.66 \pm 0.54$  & \ldots &\ldots  \\
1993 May 12.85--May 13.48 & 46.05--46.68  & ROSAT-HRI &
$12.47 \pm 1.07$  & \ldots & \ldots \\
1993 May 18.87--May 19.86 & 52.07--53.06  & ASCA &
$ 13.88 \pm 6.15 $  & $29.16 \pm 3.97$ & $ 50.15 \pm 14.04 $ \\
1993 Oct 24.60--Oct 25.59 & 203.30--204.29 & ASCA & \ldots  & $3.84 \pm 2.74$ & \ldots \\
1993 Nov 01.68 & 211.38 & ROSAT-PSPC & $7.47 \pm 0.23$  & \ldots & \ldots \\
1993 Nov 07.90 & 217.60 & ROSAT-PSPC & $8.08 \pm 0.50$  &\ldots  &\ldots  \\
1994 Apr 01.45 & 362.15 & ROSAT-PSPC & $3.59 \pm 0.52$ &\ldots  &\ldots  \\
1994 Apr 01.68 & 362.38 & ASCA & $8.83 \pm 3.15 $  & $2.30 \pm 2.06$ & $11.51 \pm 5.36$  \\
1994 Oct 19.17--Oct 21.31 & 562.82--564.96  & ROSAT-HRI &
$4.35 \pm 0.36$ &\ldots  &\ldots  \\
1994 Oct 21.18--Oct 22.33 & 564.83--565.98  & ASCA &
$4.89 \pm 2.37$  & $<0.47$&
$ 4.89 \pm 1.89 $  \\
1995 Apr 13.71--May 04.90 & 739.36--763.70 & ROSAT-HRI &
$5.35 \pm 0.43$  &\ldots  &\ldots  \\
1995 Oct 19.00 & 909.65 & ROSAT-HRI & $6.74 \pm 0.48$ &\ldots  &\ldots  \\
1996 Apr 15.88--May 07.83 & 1106.53--1144.48 & ROSAT-HRI &
$6.59 \pm 0.48$  & \ldots &\ldots  \\
1996 Nov 04.50 & 1309.15 & ROSAT-HRI & $5.54 \pm 0.80$ & \ldots &\ldots  \\
1997 Mar 31.10 & 1455.75 & ROSAT-HRI & $6.26 \pm 0.46$  & \ldots &\ldots  \\
1997 Sep 30.44--Oct 16.00& 1639.09--1654.65 & ROSAT-HRI &
$5.83 \pm 0.43$  & \ldots &\ldots  \\
1998 Mar 26.10 & 1815.75 & ROSAT-HRI & $4.08 \pm 0.50$ & \ldots &\ldots  \\
2000 Mar 21.03 & 2541.68 & {\it Chandra} ACIS-S
& $ 4.44 \pm 1.40 $ & $ 1.75 \pm 0.44 $ & $ 5.89 \pm 1.69$ \\
2000 May 07.23 & 2588.88 & {\it Chandra} ACIS-S
& $ 3.88\pm 0.14$ & $ 1.66\pm 0.18$ &  $ 5.26\pm 0.40$\\
2001 Apr 22.32 & 2938.97 & XMM-Newton EPIC-PN
& $4.54 \pm 0.25$ & $1.25 \pm0.24 $ &  $ 5.55 \pm 0.44 $\\
2005 Feb 24.29--Aug 14.41\tablenotemark{a} & 4341.94--4513.06 & {\it Chandra} ACIS-S (HETG)
& $ 1.77\pm 0.18$ & $ 0.43\pm0.08 $ &  $2.10 \pm 0.20$\\
2005 Apr 21.00--Aug 25.05\tablenotemark{b} & 4397.65--4523.70 & {\it Swift}-XRT
& $ < 2.24$ & $<0.41$ &  $<2.46$\\
2005 May 26.16--Jun 15.07\tablenotemark{c} & 4432.81--4452.72 & {\it Chandra} ACIS-S
& $ 2.16 \pm 0.11$ & $0.46 \pm 0.02$ &  $2.53 \pm 0.13$\\
2005 Jun 18.48--Jul 06.33\tablenotemark{d} & 4456.13--4473.98 & {\it Chandra} ACIS-S
& $ 2.24 \pm 0.38$ & $0.49 \pm 0.10$ &  $ 2.63 \pm 0.44$\\
2006 Feb 08.85--Aug 12.68\tablenotemark{e} & 4691.5--4876.33 & {\it Chandra} ACIS-S(HETG)
& $ 1.61\pm 0.60$ & $ 0.24\pm0.07 $ &  $1.78 \pm 0.67$\\
2006 Jun 24.00--Nov 18.04\tablenotemark{f} & 4826.65-4973.69 & {\it Swift}-XRT
 & $ 1.91\pm 0.32$ & $0.33 \pm0.05 $ &  $ 2.08\pm 0.35$\\
2008 Jan 09.04--15.47\tablenotemark{g} & 5385.69-5392.12 & {\it Swift}-XRT
 & $ 2.02\pm 0.55$ & $ 0.37\pm 0.08$ &  $ 2.22\pm 0.59$\\
2008 Feb 01.00 & 5407.65 & {\it Chandra} ACIS-S
& $ 1.67\pm 0.1 $ & $ 0.30\pm 0.08$ &  $ 1.85\pm 0.1$\\
 \enddata
 \tablenotetext{a}{Combined observations with IDs 5600, 5601, 6174, 6346 and 6347}
 \tablenotetext{b}{Combined observations with IDs 00035059001 and 00035059002}
 \tablenotetext{c}{Combined observations with IDs 5935, 5936, 5937, 5939,
  5940 and 5941}
  \tablenotetext{d}{Combined observations with IDs 5943, 5944, 5945, 5946,
  5947, 5948 and 5949}
 \tablenotetext{e}{Combined observations with IDs 6892--6901}
 \tablenotetext{f}{Combined observations with IDs 00035059003  and 00035059004}
 \tablenotetext{g}{Combined observations with IDs 00036557001, 00036557002
and 00036557003}
 \end{deluxetable}

\clearpage

\begin{landscape}
\begin{deluxetable}{l|l|lll|lll|lll}
\tabletypesize{\footnotesize}
\tablecaption{Timescales in days for reverse shock to remain radiative
for various values of ejecta velocity, mass loss rate, composition and
density index \tablenotemark{a}
\label{tab:tcool}}
\tablewidth{0pt}
\tablehead{
\colhead{Composition} & \colhead{Density} & \multicolumn{3}{c}{$V_4=0.5$} &
\multicolumn{3}{c}{$V_4=1.0$} & \multicolumn{3}{c}{$V_4=2.0$} \\
\colhead{} & \colhead{index ($n$)} & \colhead{$\dot M_{-5}=1$}
& \colhead{$\dot M_{-5}=4$} & \colhead{$\dot M_{-5}=10$}
& \colhead{$\dot M_{-5}=1$} & \colhead{$\dot M_{-5}=4$}
& \colhead{$\dot M_{-5}=10$} & \colhead{$\dot M_{-5}=1$} &
\colhead{$\dot M_{-5}=4$} & \colhead{$\dot M_{-5}=10$}
}
\startdata
Solar &7  &60 & 250&570 &4 & 16&40 &0.4 & 1.6& 4 \\
      &12 &3080 & $1.2\times10^4$&$3\times10^5$ &160 &630 &1580 &8 &32 & 80 \\
      &20 &$1.3\times10^5$ &$5.2\times10^5$ &$1.3\times10^6$ &3000
&$1.2\times10^4$ &$2.9\times10^4$ &125 &500 &1250  \\
Helium &7  &190 & 760&1900 &19 &76 &190 &2 &8 &20  \\
      &12 & $1.2\times10^4$& $4.9\times10^4$& $1.2\times10^5$&473 &1900
 & 4730&40 &160 & 400 \\
      &20 & $4.4\times10^5$&$1.8\times10^6$ & $4.4\times10^6$
&$1.2\times10^4$ &$4.9\times10^4$ &$1.2\times10^5$  &400 &1700 & 4000 \\
Oxygen &7  &6310 & $2.5\times10^4$&$6.3\times10^4$ &470 &
1900&4700 &42 & 170&420  \\
      &12 & $5.5\times10^5$& $2.2\times10^6$ & $5.5\times10^6$
&$1.7\times10^4$ &$6.6\times10^4$ & $1.7\times10^5$&977 &3900 &9770  \\
      &20 &$3\times10^7$ &$1.2\times10^8$ &$3\times10^8$ &$5.5\times10^5$ &
$2.2\times10^6$ &
$5.5\times10^6$ &$1.5\times10^4$ & $5.9\times10^4$& $1.5\times10^5$ \\
\enddata
\tablenotetext{a}{In this Table $V_4$ is the ejecta velocity in
$10^4$ km s$^{-1}$, and $\dot M_{-5}$ is mass loss rate in units of
$10^{-5}\, \rm M_\odot\, yr^{-1}$}
\end{deluxetable}
\end{landscape}

\clearpage

\begin{deluxetable}{lllr}
\tabletypesize{\small}
\tablecaption{H-$\alpha$ luminosities for SN 1993J
\label{tab:halpha}}
\tablewidth{0pt}
\tablehead{
\colhead{Days since} &
 \colhead{Luminosity} & \colhead{Observatory} & \colhead{Reference}\\
\colhead{explosion}  & \colhead{[erg/s]} & \colhead{} & \colhead{}
}
\startdata
171 &  $ 14.98 \times 10^{ 38 }$ & Asiago & \citet{pat95}\\
205 &  $ 8.67 \times 10^{ 38 }$ & Asiago & \citet{pat95}\\
236 &  $ 4.96 \times 10^{ 38 }$ & Asiago & \citet{pat95}\\
255 &  $ 4.11 \times 10^{ 38 }$ & Asiago & \citet{pat95}\\
299 &  $ 2.29 \times 10^{ 38 }$ & Asiago & \citet{pat95}\\
367 &  $ 1.28 \times 10^{ 38 }$ & Asiago & \citet{pat95}\\
553 & $ 2.50 \times 10^{ 38 }$ & Lick & \cite{mat00b} \\
670 & $ 3.00 \times 10^{ 38 }$ & Keck & \cite{mat00b} \\
687 & $ 2.88 \times 10^{ 38 }$ & NAOC & \cite{zha04} \\
700 & $ 2.75 \times 10^{ 38 }$ & NAOC & \cite{zha04} \\
881 & $ 2.50 \times 10^{ 38 }$ & Lick & \cite{mat00b} \\
976 & $ 4.80 \times 10^{ 38 }$ & Keck & \cite{mat00}\\
986 & $ 2.19 \times 10^{ 38 }$ & NAOC & \cite{zha04} \\
998 & $ 1.95 \times 10^{ 38 }$ & NAOC & \cite{zha04} \\
1034 & $ 1.99 \times 10^{ 38 }$ & NAOC & \cite{zha04} \\
1280 & $ 1.74 \times 10^{ 38 }$ & NAOC & \cite{zha04} \\
1318 & $ 1.29 \times 10^{ 38 }$ & NAOC & \cite{zha04} \\
1395 & $ 1.38 \times 10^{ 38 }$ & NAOC & \cite{zha04} \\
1729 & $ 1.10 \times 10^{ 38 }$ & NAOC & \cite{zha04} \\
1766 &  $ 0.86 \times 10^{ 38 }$ & Keck & \cite{mat00b} \\
2066 & $ 0.65 \times 10^{ 38 }$ & NAOC & \cite{zha04} \\
2454  & $ 0.37 \times 10^{ 38 }$ & Keck & \cite{mat00b}\\
3149 & $ 0.50 \times 10^{ 38 }$ & NAOC & \cite{zha04} \\
3401 & $ 0.43 \times 10^{ 38 }$ & NAOC & \cite{zha04} \\
3503 & $ 0.43 \times 10^{ 38 }$ & NAOC & \cite{zha04} \\
3610  & $ 0.28 \times 10^{ 38 }$ & Keck & \cite{fil05} \\
\enddata
\end{deluxetable}

\clearpage

\clearpage

\clearpage

\clearpage

\clearpage

\end{document}